\documentclass{an}
\usepackage{graphicx}
\usepackage{times}
\Volume{0}               
\Year{2004}              
\Pagespan{000}{000}      

\begin{document}
\lhead[\thepage]{D.Steeghs: Doppler Tomography}
\rhead[Astron. Nachr./AN~{\bf 325}, No. \volume\ (\yearofpublic)]{\thepage}
\headnote{Astron. Nachr./AN {\bf 325}, No. \volume, \pages\ (\yearofpublic) /
{\bf DOI} 10.1002/asna.\yearofpublic1XXXX}

\title{Doppler Tomography of accretion in binaries}

\author{D.Steeghs}
\institute{Harvard-Smithsonian Center for Astrophysics}
\date{Received {date}; 
accepted {date};
published online {date}} 

\abstract{ Since its conception, Doppler tomography has matured into a versatile and
widely used tool. It exploits the information contained in the
highly-structured spectral line-profiles typically observed in
mass-transferring binaries. Using inversion techniques akin to medical
imaging, it permits the reconstruction of Doppler
 maps that image the accretion flow on micro-arcsecond scales.  I
summarise the basic concepts behind the technique and highlight two recent
results; the use of donor star emission as a means to system parameter
determination, and the real-time movies of the evolving accretion flow in
the cataclysmic variable WZ Sge during its 2001 outburst. I conclude with
future opportunities in Doppler tomography by exploiting the combination
of superior data sets, second generation reconstruction codes and
simulated theoretical tomograms to delve deeper into the physics of
accretion flows. 
}

\correspondence{dsteeghs@cfa.harvard.edu\\ For movies: Follow the links on
the publishers web-pages.\\
http://www3.interscience.wiley.com/cgi-bin/jtoc?ID=60500255\\ as well as
on the web-pages of the editorial office:\\ www.aip.de/AN/movies/ }

\maketitle

\section{Introduction}

The detection of broad and strong emission lines is one of the
hallmarks of an accreting system. The luminous supersonic accretion
flows in cataclysmic variables (CVs) and X-ray binaries (XRBs) lead to complex and
highly time-dependent line profiles across the electromagnetic
spectrum. The interpretation of such broad lines in the spectrum of
the CV AE Aqr as due to Doppler motions in a binary system (Crawford
\& Kraft 1956) formed the foundation for the development of our
standard accretion scenario involving mass transfer via Roche-lobe overflow.
The possibility of using these line
profiles as a powerful diagnostic of the accretion flow was also recognized
early on.  For example, Greenstein \& Kraft (1959) analysed the nature of 
emission line eclipses and Smak (1969) calculated synthetic line profiles
formed in an extended accretion disc around the accretor. The time-dependent velocity structure of emission lines thus quickly became an
important tool.
Doppler tomography was developed in the late 1980's by Keith Horne
and Tom Marsh, who aimed to exploit  the kinematical information
contained in the observed line profiles and recover a
{\it model-independent} map that
spatially resolves the distribution of line emission in the binary (Marsh \& Horne 1988). It was
recognized that the observed line profiles at each orbital phase
provide a projection of the accretion flow along the line of
sight. Given sufficient observed projections, these profiles can then be
inverted into a 2D image, very much like the CAT-scanning procedures used in medical
imaging. An excellent recent review of Doppler tomography can be
found in Marsh (2001). This concise review summarises the key ingredients of
Doppler tomography and focuses on future opportunities
for extending Doppler mapping methods in the light of enhanced
capabilities with large aperture telescopes. The growing popularity of
Doppler tomography is clearly illustrated by the large number of
applications of the technique that can be found in the recent
literature as well as in several other contributions to this volume.

\section{From data to Doppler map}

With Doppler tomography, a 2D data set consisting of a time series of
line profiles is inverted into a 2D Doppler tomogram. Since the
dataset provides us with projected radial velocities of the emitting
gas, the Doppler tomogram one reconstructs provides the distribution
of the line emission in the binary in a {\it velocity} coordinate
frame. This permits image reconstructions without detailed \`{a} priori assumptions
concerning the nature of the flow that one is mapping. The velocity
coordinate frame greatly
increases the flexibility of Doppler tomography and significantly
simplifies the inversion process compared to image reconstructions
in the Cartesian X-Y frame.  Nonetheless, some implicit assumptions
are made when performing Doppler tomography, and these must be borne
in mind when interpreting reconstructed accretion flow maps (Marsh 2001;
Steeghs 2003).
The rather unfamiliar velocity coordinate frame of Doppler tomograms
is the main hurdle a novice user of Doppler tomography has to
overcome. In the Doppler tomogram coordinate frame, each line source
is characterised by its inertial velocity vector in the orbital plane, {\bf V}=$(V_x,V_y)$, where
the binary center of mass is at the origin, the $x$-axis points from the
accretor to the donor and the $y$-axis points in the direction of motion
of the donor, as viewed in the co-rotating frame of the binary. Each
source then traces a sinusoidal radial velocity curve as a function of
the orbital phase ($\phi$) centered on the systemic velocity of the
binary ($\gamma$);
\[
V(\phi) = \gamma - V_x \cos{2\pi\phi} + V_y\sin{2\pi\phi}. \]
The observed line profiles are the
projection of the radial velocities and intensities of all velocity
vectors considered;
\[
F(v_r,\phi) = \int I(V_x,V_y) \star g(V-v_r) \mbox{ } dV_xdV_y , \]
with $g(V-v_r)$ describing the local line profile intensity at a Doppler
shift of $V-v_r$ and $I(V_x,V_y)$ the image value of the Doppler map
at the corresponding velocity grid point. The inversion from data-set to Doppler map can be
achieved via either a Radon transform in the form of filtered
back-projection (Horne 1991) or via regularised fitting using maximum entropy or
a similar quantity as the regularising function (Marsh \& Horne
1988).  The superior handling of noise and image artifacts with
maximum entropy mapping, combined with its flexibility and usage of a
formal goodness of fit parameter in the form of $\chi^2$ make it the preferable
inversion choice in most cases.

Despite the unfamiliar coordinate frame of Doppler tomograms,
it is relatively straightforward to identify and resolve prominent
emission sites such as the extended accretion disc, the gas stream and
its impact on the outer disc (Fig. 1), the magnetically channeled flow in
polars and emission from and around the mass donor star. Spatial
asymmetries in the emissivity of the accretion disc map into similar
structures in the inside-out Doppler projection of that disc. Since each of these sites has a completely different
position-velocity relation, it would be impossible to reconstruct the
equivalent position image without imposing detailed assumptions
concerning the (unknown) translation between velocity and
position. Therefore it is advised to compare data with models by
presenting models in the velocity coordinate frame, as opposed to
trying to reconstruct the data in an XY-frame. Then a
direct and quantitative comparison between model and data can be
made. Thus Doppler tomograms in the literature  are now not only
generated from observed line profiles, but also from theoretical simulations (e.g Wynn, King \& Horne 1997).

\section{Tomography highlights}

I refer to the several other contributions to this volume as well as the proceedings of the 2000 Brussels tomography
workshop (Boffin, Steeghs \& Cuypers 2001) for specific applications
of Doppler tomography of accreting binaries.  In disc accreting
systems, tomograms have demonstrated that in many cases accretion disc
flows are highly asymmetric. The interaction between the in-falling
stream and the outer edge of the disc manifests itself in a prominent
bright spot as well as more extended disc asymmetries downstream
(e.g. Figure 1). With
Doppler maps, we have the prospect of studying this interaction in
detail using a range of different lines. The disc itself generates
strong asymmetries most notably the two armed tidal spiral structures
observed in dwarf novae during outburst (Morales-Rueda, this volume). 
Doppler tomograms have also been instrumental in the mapping of the
magnetically controlled flows in polars (Schwope, this volume) and in
highlighting unusual emission-line kinematics in high-mass transfer CVs and
some X-ray binaries. Below, I discuss two recent results I have been
personally involved with in more detail.

\subsection{Donor star emission and system parameters}
  Apart
from imaging the geometry and dynamics of the accretion flow in a
wide range of settings, Doppler tomography can also be used as a means
toward the determination of basic system parameters.  In particular,
a contribution from the donor star is commonly observed in high-mass
transfer rate CVs, polars and X-ray binaries. Whereas a radial
velocity analysis of photospheric absorption features is the common
technique for measuring the orbital velocity of the donor, there are
many cases where such photospheric features are too weak to be
detected. This has severely hampered our ability to extract basic
system parameters for such binaries.  In those cases, emission
components from the irradiated donor are the only means of
establishing its radial velocity and thereby constrain its mass. Since donor star emission
typically contributes only a modest fraction to the overall emission
line flux which tends to be dominated by the extended accretion flow,
traditional radial velocity analysis of such features is
problematic. On the other hand, a Doppler tomogram cleanly separates
such donor emission from disc emission since emission from the Roche
lobe-shaped donor stars maps onto a Roche lobe-shaped region along the
positive $V_y$ axis in Doppler tomograms (Figs 1 \& 2). This has the added advantage
that the Doppler tomogram makes use of all the observed profiles at
once and can thus identify very weak donor contributions that are too
feeble to be identified in the individual line profiles. The main
disadvantage of this method is that the emission traces only that
part of the donor that is exposed to the ionising radiation from the
hot accretion flow, and thus is biased toward the front side of the
donor. A correction has to be made therefore to estimate the true
orbital velocity of the donor's center of mass as is required for
system parameter estimates. Similar, but opposite, biases are present in the radial
velocities curves derived from photospheric absorption line, which
tend to over-estimate the amplitude since the absorption is weighted
towards the back side of the donor. In
some cases, the presence of both photospheric absorption as well as
donor star emission lines allow us to investigate these relative
biases (e.g. Schwope, this volume). 3D mapping of the distribution
of donor star emission/absorption can be achieved via Roche tomography
(Watson, this volume).
Nonetheless, despite these
biases, emission line radial velocity measurements of donor stars
allow us to derive firm limits on the radial velocity amplitude of the
donor star in systems for which no other dynamical constraint is
available.  In dwarf novae during outburst, the donor star is commonly
present. This finally gave us a view of
the hitherto unseen low mass donor star in WZ Sge during the 2001
outburst, when the hot accretion flow irradiated its Roche lobe (Figure 2; Steeghs et al. 2001).
The detection of sharp donor
emission components in the Bowen blend in the prototypical XRB Sco X-1
(Steeghs \& Casares, 2002) illustrates the prospect for extending
this technique to a significant number of neutron star (Casares et
al. 2003) and black hole binaries (Hynes et al. 2003) as well.

\begin{figure}
\resizebox{\hsize}{!}
{\includegraphics[]{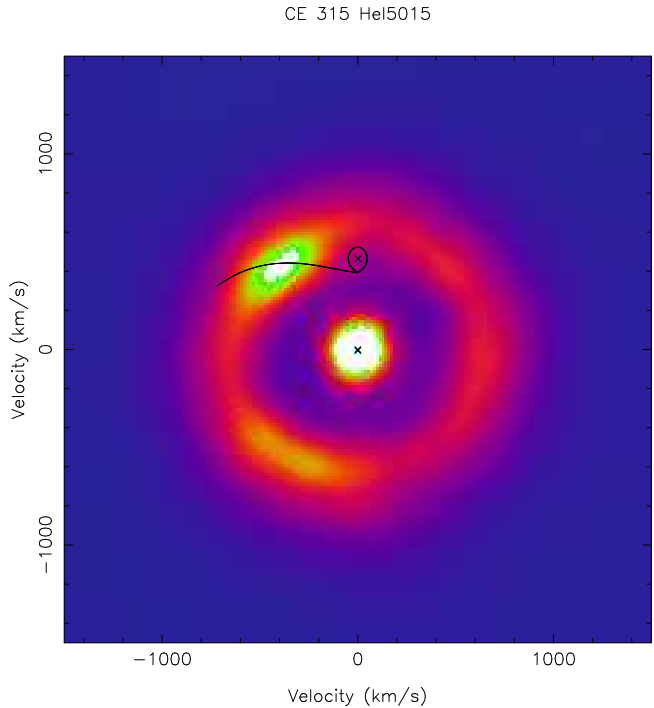}}
\caption{HeI 5015\AA~Doppler tomogram of the hydrogen deficient binary CE-315. The cross near the system's center
of mass denotes the location of the accreting white dwarf, a
uncharacteristically strong emission-line source. The donor star
together with a ballistic gas stream trajectory is over plotted. A
prominent bright spot is visible at the stream-disc impact
location. The system has an extreme mass ratio of $q=0.0125$ and a
very low mass donor star of order 5 Jupiter masses due to gigayears of
mass transfer (Steeghs, Nelemans \& Marsh, in press).}
\label{ce315}
\end{figure}

\begin{figure}
\resizebox{\hsize}{!}
{\includegraphics[]{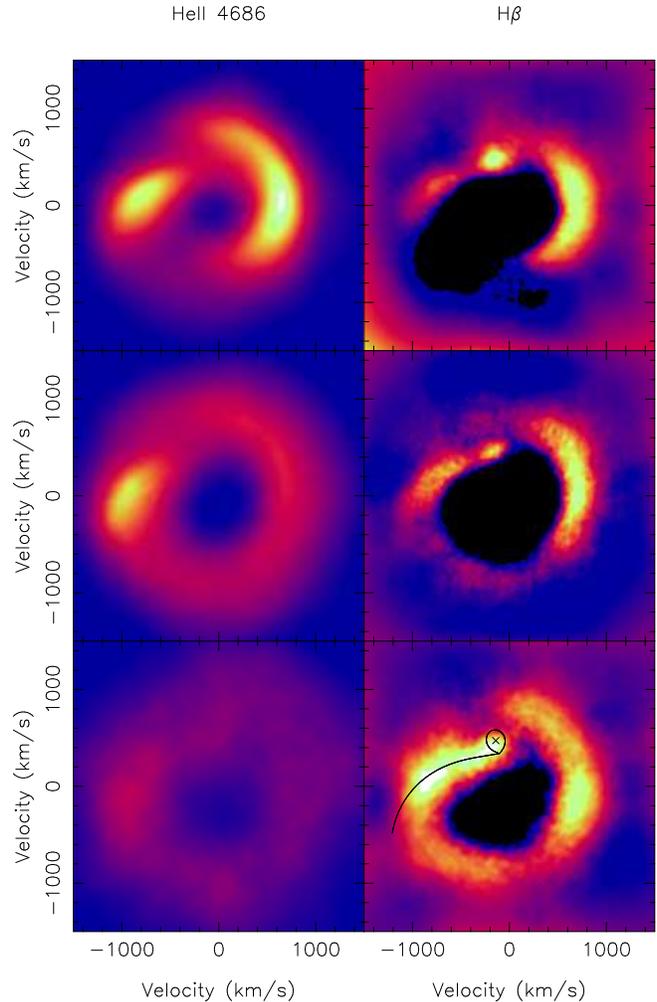}}
\caption{WZ Sge outburst Doppler tomograms of the HeII 4686 (left) and
$H\beta$ emission lines. Top panels are obtained 3
days into the outburst and show a hot distorted disc, strong HeII
emission and two armed spirals. The middle panels are at day 9 and the bottom
panels at day 19. Note the persistent presence of a compact emission source
in the $H\beta$ maps  attributed to the donor star, as indicated in
the bottom-right panel.}
\label{figlabel}
\end{figure}

\subsection{Time-lapsed tomography: WZ Sge in outburst}

Doppler tomograms provide snapshots of the accretion geometry in a
wide range of systems. In many cases, data sets covering several orbits
are averaged together to provide a time-averaged image of the accretion
flow in the co-rotating frame. Given that in principle only half an
orbit is required for the reconstruction of a well constrained
tomogram, a sequence of Doppler maps can track the evolution of
systems on timescales of order their orbital period and longer. Armed
with a suitably long data-set we can thus construct real-time movies of
the changing accretion dynamics in binary systems. The dwarf novae and
X-ray transients are excellent targets for such time-lapsed tomography
experiments, since their recurrent outbursts are characterised by
timescales of days to weeks, and can thus be probed by an intensive
observing campaign.  The 2001 outburst of the short-period CV WZ Sge led to
such an extensive campaign with a large number of ground-based as well
as space-based observatories studying this rare outburst. With
over 35 nights of phase-resolved spectroscopy, the WZ Sge outburst
provided a unique opportunity for time-lapsed tomography.
During the first weeks of the outburst, which started on July
23rd 2001, time-resolved spectroscopic data obtained using the INT and
WHT telescopes on La Palma provided over 10 nights of data with sufficient resolution
for tomography. This data allowed the calculation of emission-line
movies displaying the evolving accretion flow in the form of a
sequence of Doppler tomograms covering the first 20 days of the outburst. In Figure 2, I display some
representative frames of these movies. The movies themselves are
available via the online edition of this journal. Below I describe the
key features of these movies.

During the first few days of the outburst, the large, hot accretion disc is distorted by a two armed  spiral structure,
very much like the structures seen in IP Peg and U Gem (top panels
in Figure 2, Steeghs 2001, Groot 2001). This disc
asymmetry weakens as the HeII line flux decays, with the arm closest
to the gas stream impact site persisting longer (middle-left panel). Roughly 14 days into the outburst, a short re-brightening occurs
which corresponds to an increase in the HeII flux and the
renewed presence of both arms. These quickly dissipate as the disc
cools down, until the HeII line becomes undetectable (bottom-left panel). The
Balmer emission suffers from significant self-absorption in the early
stages, and makes a slow transition to a pure emission line.  The
variable presence of an emission component from the mass donor is also
apparent. The most intriguing
development is  seen around the expected location of the bright
spot. After the short re-brightening that was seen in the HeII line,
the H$\beta$ line shows an extended emission structure bridging the
mass donor and the outer disc; i.e. along the location of the gas stream (bottom-right panel). More extensive
analysis is in progress, but it has all
the hallmarks of a short-lived burst of enhanced mass transfer from
the mass donor. Such a mass transfer burst has been invoked in the
past to explain the unusual nature of the WZ Sge outbursts, and  extensive photometry by Patterson et al. (2002) indicates the presence of
an enhanced bright spot signature in the orbital light-curves around
the same time.
These first-generation Doppler movies illustrate the potential of
time-lapsed tomography experiments. They warrant more efforts to
secure similar data sets and provide detailed real-time evolution of the accretion flow. The outburst timescales of dwarf novae provide 
unique opportunities to scrutinize our understanding of disc
instabilities and the physics of angular momentum transport.

\section{The road ahead}

Since its conception, Doppler tomography has evolved into a popular
tool for analysing the complex line profiles from a wide range of
accreting binaries. With the advent of large telescopes and a new
generation of spectrographs, it is now not only possible to apply
tomography to a wider range of systems by going fainter, but also push
the technique further by acquiring high-quality data sets (in terms of
superior time and wavelength resolution and signal to noise) of
relatively bright systems.
With the quality and depth of data sets improving, it is also
important to evolve the tomography codes and take full advantage of
the information contained in such data sets. 
Although the number of assumptions that go into Doppler tomography are
limited, there are prospects for relaxing these assumptions and
improving the flexibility and capabilities of tomography further.  I explored one such
extension recently by relaxing the assumption that the assumed flux
from any point in the binary is constant in time (Steeghs 2003). This
is one of the standard assumptions that is clearly violated in many
cases. By allowing the flux to vary harmonically as a function of the
orbital phase, the modulated Doppler tomography implementation can
describe complex phase variability in the lines. The first
applications include mapping the bright spot region in GP Com
(Morales-Rueda et al. 2003) and the anisotropic emission from the
spiral arms in IP Pegasi (Steeghs 2003). 
For the magnetically funneled flows in polars and intermediate polars
it is clear that motions out of the orbital plane are significant,
whereas Doppler tomography strictly maps the $V_x$ and $V_y$
components. This has led to a number of mapping experiments where the
3D geometry of the flow was prescribed before mapping the brightness
distribution across it. However, in principle, the $V_z$ component can
be incorporated in a 3D version of Doppler tomography, whereby the
image space is now a cube, each slice corresponding to the
distribution in the X-Y plane for a given $V_z$. Tom Marsh has
started some exploratory work in this direction (Marsh, private
communication).

In order to extract physical parameters from the flow, the information
contained in different emission lines can be combined. Rather than
mapping each emission line on its own, images can be reconstructed in
terms of physical parameters. By analogy with the physical parameter
eclipse mapping technique (Vrielmann, Hessman \& Horne 1999), a simple LTE atmosphere
model characterised by temperature and density can be used to recover
density and temperature Doppler maps, constrained by a data set
covering a sufficiently diverse set of lines. Such physical parameter
constraints would be invaluable.
Doppler tomography maps accretion flows in a wide range of settings on
micro-arcsecond scales. It continues to grow in popularity and is approaching
the point of becoming a standard tool for emission line analysis of
CVs and X-ray binaries.  A combination of better data sets, more
capable reconstruction codes and theoretical tomograms provides
plenty of opportunities for novel Doppler tomography applications in
the years ahead.

\end{document}